# A METHOD FOR STABILITY ANALYSIS OF MAGNETIC BEARINGS : BASIC STABILITY CRITERIA

**B Shayak**

Department of Theoretical and Applied Mechanics,
School of Mechanical and Aerospace Engineering,
Cornell University,
Ithaca – 14853,
New York, USA

sb2344@cornell.edu , shayak.2015@iitkalumni.org

\*

## Classification

## Abstract

In this work I outline a general procedure for dynamic modeling and stability analysis of a magnetic bearing, which is a rotating shaft confined inside a chamber through electromagnetic forces alone. I consider the simplest type of self-propelled bearing, namely a permanent magnet synchronous motor and an induction motor rotor freely suspended inside the corresponding stator, and having no eccentricity-fedback control algorithm. Writing Euler's equations for the rotor mechanics and Maxwell's equations for the electromagnetic field leads to a systematic technique for analysing the dynamics of the complete system. Physical arguments indicate that that two essential components for rotor confinement are a spatial gradient in the stator magnetic field and a torque angle lying in the second quadrant. These predictions are confirmed through the linear stability analysis. The direct practical utility of the results is mitigated by the presence of a repeated eigenvalue in the linearized equations. Despite this limitation, the analysis presented can act as a good starting point for more accurate treatments of other magnetic bearing configurations.

\*   \*   \*   \*   \*

**Introduction**

The problem of designing a magnetic bearing which can combine useful and desirable features such as simplicity, robustness, high load support capacity, high efficiency, and above all, high intrinsic stability, is one which is now at the forefront of electromechanical engineering research. Although a miscellany of designs exist, none of them has the combination of qualities required to go beyond speciality applications into the world of everyday appliances and gadgets. One of the reasons behind this is the absence of an universal analytic method which can yield the stability and performance characteristics of any given magnetic bearing configuration. An early work to consider modeling of non-contact bearings is by K A CONNOR and J A TICHY [1]; their treatment is primarily heuristic. A refinement has been presented by RICHARD POST and D D RYUTOV [2] who compute the equivalent stiffness of the rotor confining springs for a specific electromagnetic configuration (Halbach array). A theoretical analysis valid for any arbitrary electromagnetic configuration has been initiated by ALEXEI FILATOV et. al. [3] who discuss several general principles behind the bearings' operation. TORBJORN LEMBKE [4] has separately considered lumped parameter equivalents of the electromagnetic and mechanical modules. The rotor has been treated as a Jeffcott rotor and some modes of motion (rotation, whirling) have been examined, however a discussion of stability is lacking here. A similar model has been considered by VIRGINIE KLUYSKENS et. al. [5]. NICOLA AMATI et. al. [6] have performed a more detailed study of this variety of model; even so they employ many restrictive simplifications such as the unquestioned existence of electromagnetic stiffnesses, assumption of steady state electromagnetic dynamics and absence of rotational dynamics of the rotor. The most advanced model till date is by JOAQUIM DETONI [7] who has first separately analysed the electromagnetic and mechanical modules and then combined the two to perform a stability analysis. He has paid attention to the field configuration of the system, thus going beyond the standard heuristic type model and has also used a gyroscopic rotor model as against a Jeffcott one. Nevertheless, his analysis is not fully complete in that a) the electrodynamic steady state is assumed during mechanical analysis, b) the angle of precession and nutation of the rotor are both assumed small, c) the effect of the (possibly large) angular momentum of the rotor has not been considered. Further, a limitation common to all the cited works is the absence of discussion regarding the effect of the motor used to power the shaft.

In this work I propose a technique for analysing a magnetic bearing, which accepts as input the electromagnetic configuration of the rotor and stator and yields analytical formulae for any parameter related to stability of performance of the bearing. Although such a work might appear backdated, given the advanced state of simulational methods in today's world, an analytical solution to any problem is of invaluable utility as a starting point for numerical and experimental work. In a complex problem such as the present one, the parameter space is four, five or higher dimensional and analysis is the only systematic method for exploring this vast region. Further, a theoretical calculation provides insights which are impossible to obtain by any other means, and these insights can be of enormous utility in designing new bearing configurations.

As the 'test case', I confine myself here to the simplest possible self-propelled magnetic bearing structure, which is an alternating current motor rotor freely suspended inside the corresponding stator. Both permanent magnet synchronous motor (PMSM) and induction motor (IM) rotor will be considered, and basic criteria obtained for their stable operation. The elementary nature of the configuration precludes its direct application in the industry, but we hope that the method shown here can be applied to more realistic situations with greater benefit.

# 1  The System dynamic model

This part of the calculation is lengthy, so I split it into several pieces, individually analyse each piece and then put the pieces together to form a whole. Each piece gets one Subsection.

### A.  STRUCTURE OF THE MECHANICAL EQUATION

In this Subsection I formulate the left hand sides (LHS) of the equations describing the evolution of the mechanical variables. A cylindrical three phase stator is mounted in the laboratory reference frame, which is fixed. Currents are applied to this stator so as to create a rotating magnetic field inside it. This field is quadrupolar or higher multipolar, a condition necessary to achieve trapping of an electromagnetic body inside it (see the next Subsection). A Cartesian coordinate basis $x,y,z$ is attached to the stator with the origin at its geometrical centre and $z$ along its axis of symmetry. This stator is common to both induction motor (IM) and permanent magnet synchronous motor (PMSM) models. The rotor is cylindrical with radius $r_0$ and height $2h$ and is freely suspended inside the stator cavity. In the unperturbed state it is coaxial with the stator. Three coordinates and three angles have to be used to completely specify its position and orientation. It is natural to measure the displacement of the rotor centre of mass (CM) in the stator frame itself; let the rotor CM coordinates be $x_{CM}$, $y_{CM}$ and $z_{CM}$. These readily yield the structure of the translational equations of motion :

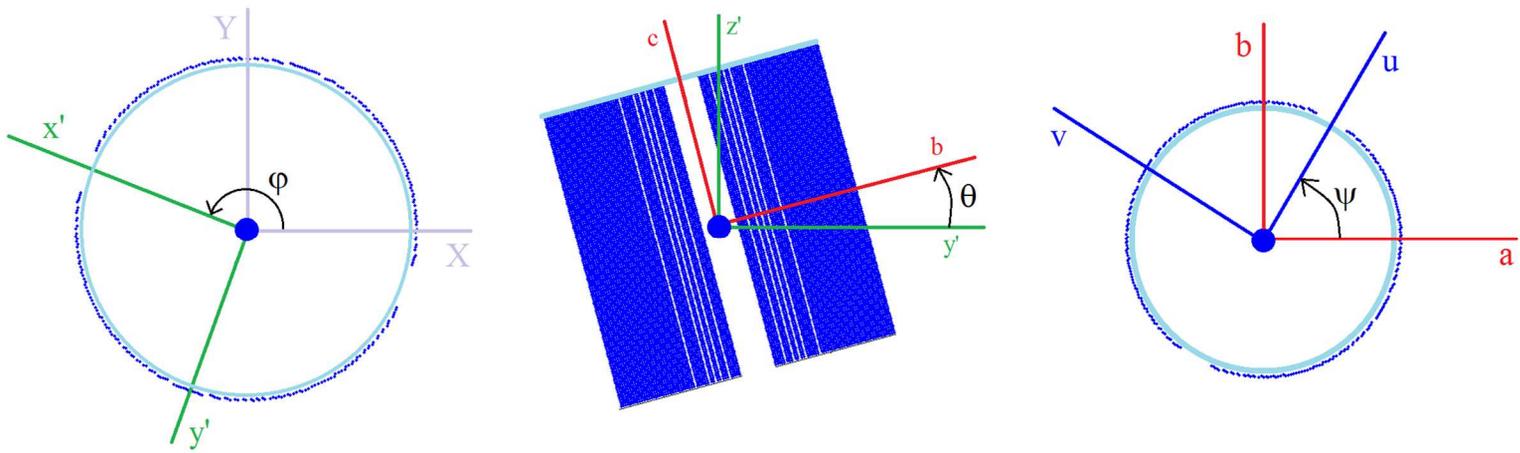

Figure 2 : *Orthographic views of the Eulerian rotations. The blue structure is the rotor, which I have depicted here as a cylindrical shell. 2L shows the view along Z axis, which comes out of the plane of the paper. A gap has been created in the rotor circumference (through which x' axis passes) to indicate a body fixed direction. Also a light blue line is used to mark the top edge of the rotor as against the bottom edge (see also Fig. 3). After the transformation, X→x', Y→y' and Z→z' although these last two axes are coincident. 2C shows the view along x' axis. After the transformation, x'→a (coincident), y'→b and z'→c. 2R shows the view along c axis. After the transformation, a→u, b→v and c→w (coincident). In all panels, only the positive halves of all axes have been shown.*

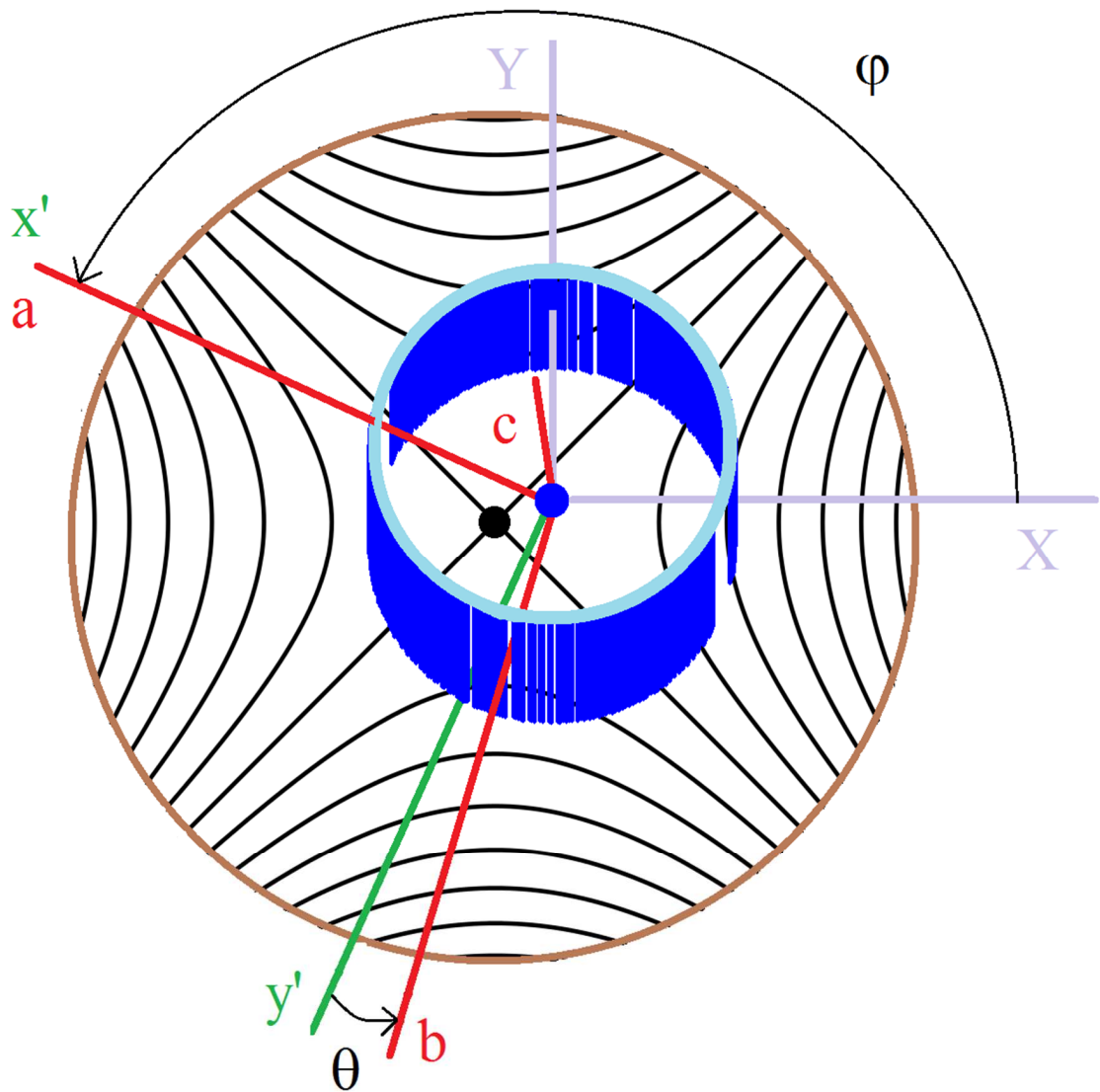

Figure 3 : *A three-dimensional view of the magnetic bearing. The large brown ring denotes the stator which produces a rotating, quadrupolar magnetic field (lines shown in black). The image of the field has been taken from Wikipedia [8]. The centre of the stator is the black dot and the x,y,z basis has its origin there. The view is along the z axis, which is positive coming out of the plane of the page. The magnetic field rotates about this axis. The rotor centre is the blue dot which is displaced from the stator centre. The gap in the circumference and the light blue top edge are useful in determining its orientation. X and Y are parallel to stator x,y but centred at the rotor CM. The other axes are all from Fig. 2. Note that the c axis is highly foreshortened as the angle of nutation is small. The u,v,w basis is not relevant for the bulk calculation so I have not shown it here.*

$$m\ddot{x}_{CM} = F_x \quad , \tag{1a}$$
$$m\ddot{y}_{CM} = F_y \quad , \tag{1b}$$
$$m\ddot{z}_{CM} = F_z \quad , \tag{1c}$$

where $m$ is the mass of the rotor and $\mathbf{F}_{xyz}$ is the force vector acting on the CM of the rotor. Of course, $\mathbf{F}$ is yet to be determined, but right now the LHS is all I am interested in. In view of future developments however I will eliminate (1c) from the system at this stage. For doability, the electromagnetic analysis will assume that the motors are very long compared to their radius, therefore, $z$ will be a redundant dimension in the calculations. Further, I will ignore the effect of gravity in this work, and assume the rotor to be force-free.

I now need to specify three angles of rotation about the CM, and I choose the Eulerian angles $\varphi, \theta, \psi$ using the x-convention of HERBERT GOLDSTEIN [9]. Letting $X,Y,Z$ denote a basis parallel to the lab $x,y,z$ but centred at the rotor CM, the following three rotations, in the order mentioned, take me from the stator orientation to the rotor orientation :

1. A rotation about $Z$ axis through an angle $\varphi$ to produce the basis $x',y',z'$. Thus $\varphi$ is the angle of precession.
2. A rotation about $x'$ axis through an angle $\theta$ to produce the basis $a,b,c$. Thus $\theta$ is the angle of nutation.
3. A rotation about $c$ axis through an angle $\psi$ to produce the basis $u,v,w$. Thus $\psi$ is the angle of spin.

Hence, $u,v,w$ is a basis which at any instant of time is aligned with a set of axes fixed permanently in the rotor. We note that this is not a genuine body-fixed frame – axis rotations can operate only between inertial or non-inertial frames but cannot transform from one to the other. Since $x,y,z$ is assumed inertial, $u,v,w$ too is so, while a frame genuinely fixed to the rotating rotor is not. I will let $u_1, v_1, w_1$ denote the non-inertial frame which is fixed in the rotor and aligned with $u,v,w$ at any instant of time. In Fig. 2 I show the three transformations occurring one after the other, while in Fig. 3 I show the effect of the combined transformation. It turns out that only one rotation matrix will be relevant later on :

$$\begin{bmatrix} \hat{\mathbf{a}} & \hat{\mathbf{b}} & \hat{\mathbf{c}} \end{bmatrix}' = \mathbf{R}_{xyz}^{abc} \begin{bmatrix} \hat{\mathbf{x}} & \hat{\mathbf{y}} & \hat{\mathbf{z}} \end{bmatrix}'$$

OR

$$\begin{bmatrix} \hat{\mathbf{a}} \\ \hat{\mathbf{b}} \\ \hat{\mathbf{c}} \end{bmatrix} = \begin{bmatrix} \cos\varphi & \sin\varphi & 0 \\ -\cos\theta\sin\varphi & \cos\theta\cos\varphi & \sin\theta \\ \sin\theta\sin\varphi & -\sin\theta\cos\varphi & \cos\theta \end{bmatrix} \begin{bmatrix} \hat{\mathbf{x}} \\ \hat{\mathbf{y}} \\ \hat{\mathbf{z}} \end{bmatrix} \quad , \tag{2}$$

for conversion from $x,y,z$ to $a,b,c$ bases and its transpose for the inverse transformation.

I need not require at this stage that the rotor be symmetric but I do assume (reasonably and accurately enough for practical purposes) that one of its principal axes coincides with its geometrical axis of symmetry, $w_1$. Then the two other axes lie in the perpendicular plane and without loss of generality I can say that the principal basis for the rotor is $u_1, v_1, w_1$. For this basis, $\boldsymbol{\omega}_{u1v1w1} = \omega_u \hat{\mathbf{u}} + \omega_v \hat{\mathbf{v}} + \omega_w \hat{\mathbf{w}}$, where $\boldsymbol{\omega}_{u1v1w1}$ is the angular velocity vector of the $u_1, v_1, w_1$ frame relative to the ground. Now, the rotor angular momentum $\mathbf{L}$ has a simple form in the $u_1, v_1, w_1$ basis ($\mathbf{L}_{u1v1w1} = I_{u1}\omega_{u1}\hat{\mathbf{u}}_1 + I_{v1}\omega_{v1}\hat{\mathbf{v}}_1 + I_{w1}\omega_{w1}\hat{\mathbf{w}}_1$) as that is principal, but the dynamics there is useless to me as the frame is non-inertial. To come out into the inertial frame, I must perform an Eulerian extraction from $u_1, v_1, w_1$ to $u,v,w$ and then equate the rate of change of angular momentum to the torque in that basis. I have

$$\frac{d\mathbf{L}_{uvw}}{dt} = \frac{d\mathbf{L}_{u1v1w1}}{dt} + \boldsymbol{\omega}_{u1v1w1} \times \mathbf{L}_{uvw} \quad , \text{ and} \tag{3a}$$

$$\frac{d\mathbf{L}_{uvw}}{dt} = \mathbf{T}_{uvw} \quad , \tag{3b}$$

where $\mathbf{T}$ denotes the torque on the rotor. Finally, due to alignment of $u,v,w$ and $u_1, v_1, w_1$ bases, the inertia tensor is the same in both; thus $\mathbf{L}_{uvw} = I_u \omega_u \hat{\mathbf{u}} + I_v \omega_v \hat{\mathbf{v}} + I_w \omega_w \hat{\mathbf{w}}$, where $I_{u1} = I_u$ and so on. This leads to the well-known form of Euler's equation

$$I_u \dot{\omega}_u + (I_w - I_v) \omega_w \omega_v = T_u \quad , \tag{4a}$$
$$I_v \dot{\omega}_v + (I_u - I_w) \omega_u \omega_w = T_v \quad , \tag{4b}$$
$$I_w \dot{\omega}_w + (I_v - I_u) \omega_v \omega_u = T_w \quad . \tag{4c}$$

The distinction between the non-inertial $u_1, v_1, w_1$ and the inertial $u,v,w$ is a subtle one and is rarely found in the literature, but it is useful for conceptual clarity.

A considerable simplification occurs however if the rotor is symmetric and I now assume that it is. (This might make me sound crazy, but I wrote (4) with a purpose, namely to show how the modeling can proceed in the absence of symmetry.) If the rotor is symmetric then a moving frame $a_1, b_1, c_1$ which at any instant of time is aligned with $a,b,c$ is a principal basis for it. Now the angular velocity of the frame $a_1,b_1,c_1$ with respect to the ground (or to $a,b,c$) is $\boldsymbol{\omega}_{a1b1c1} = \omega_a \hat{\mathbf{a}} + \omega_b \hat{\mathbf{b}}$ (note : no $\omega_c \hat{\mathbf{c}}$ !) and it is this angular velocity which must be crossed with the **L** vector while performing Eulerian extraction from $a_1,b_1,c_1$ to $a,b,c$ frame. The angular momentum is still $\mathbf{L}_{abc} = I\omega_a \hat{\mathbf{a}} + I\omega_b \hat{\mathbf{b}} + I_c \omega_c \hat{\mathbf{c}}$, where I have used $I_a=I_b=I$, and performing the extraction I have

$$I\dot{\omega}_a - I_c \omega_c \omega_b = T_a \quad , \tag{5a}$$

$$I\dot{\omega}_b + I_c \omega_a \omega_c = T_b \quad , \tag{5b}$$

$$I\dot{\omega}_c = T_c \quad . \tag{5c}$$

Prima facie there is only a cosmetic difference between (5) and (4), but that is not the case. It will turn out that the torque vector naturally evaluates in the *a,b,c* frames, and then a conversion to *u,v,w* will simply mean a lot of extra transformation and headache during the subsequent analysis. The derivation of $\boldsymbol{\omega}_{abc}$ in terms of $\dot{\theta}$, $\dot{\varphi}$ and $\dot{\psi}$ is quite simple using the definition of the rotations; the answer is

$$\boldsymbol{\omega}_{abc} = \dot{\theta}\hat{\mathbf{a}} + \dot{\varphi}\sin\theta \hat{\mathbf{b}} + (\dot{\psi} + \dot{\varphi}\cos\theta)\hat{\mathbf{c}} \quad . \tag{6}$$

Finally, it is useful at this step to factor out the air resistance which will be present in the stator cavity (in fact that is the only load which the motor will have to overcome). The simplest form of air resistance is a term proportional to linear velocity in (1) and proportional to angular velocity in (5); letting the constants be $\kappa_1$ and $\kappa_2$ I have the structure of the mechanical equation of motion as

$$m\ddot{x}_{CM} + \kappa_1 \dot{x}_{CM} = F_x \quad , \tag{7a}$$

$$m\ddot{y}_{CM} + \kappa_1 \dot{y}_{CM} = F_y \quad , \tag{7b}$$

$$I\dot{\omega}_a + \kappa_2 \omega_a - I_c \omega_c \omega_b = T_a \quad , \tag{7c}$$

$$I\dot{\omega}_b + \kappa_2 \omega_b + I_c \omega_a \omega_c = T_b \quad , \tag{7d}$$

$$I_c \dot{\omega}_c + \kappa_2 \omega_c = T_c \quad . \tag{7e}$$

With air resistance out of the equation, the only terms in **F** and **T** will be the ones arising from the electromagnetic interaction between rotor and stator. This completes the structure of the mechanical equation and gets us smoothly started on the path towards the dynamic model of the system.

### B. STRUCTURE OF THE ELECTROMAGNETIC INTERACTIONS

In the motor model I will use, I will assume that the rotor and stator carry *surface currents parallel to their axes of symmetry* in their external and internal periphery respectively. This assumption reduces the electromagnetism to an analytically doable form, and is also implicit in the widely used equivalent circuit model of a motor (which I will not touch here). I will further assume that the stator and rotor are very (infinitely) long compared to their radii. Focusing on the stator, its surface current points in the *z* direction and is a function of $\vartheta$, where $\vartheta$ is an angle in the *x,y* plane with the *x*-axis as a reference line. The most general form of this function is

$$\mathbf{K}_s = \hat{\mathbf{z}} \left[ K_{sdc} + \sum_{n=1}^{\infty} \left( K_{s1n} \cos n\vartheta + K_{s2n} \sin n\vartheta \right) \right] \quad . \tag{8}$$

Equation (8) is of course a Fourier series in $\vartheta$; the first term is the dc component, which is the total current (here and henceforth, 'surface' is implicit) flowing in the stator. For any practical stator supplied with alternating current this component is zero, so I neglect it in the subsequent analysis. The cosine and sine terms for each *n* are assigned subscripts 1 and 2. The vector potentials and magnetic fields created by this $\mathbf{K}_s$ all contain the same angular harmonics as $\mathbf{K}_s$ itself. The number of harmonics in $\mathbf{K}_s$ and their relative amplitudes are determined by the arrangement of the stator windings. I now assume that this arrangement is such that there is only one harmonic present in $\mathbf{K}_s$ i.e. $K_{s1n}$ and $K_{s2n}$ exist for only one particular $n=n_0$ and are zero for all other *n*. This assumption is made in nearly all treatments of electric motors – it generally works well because multiple harmonics cause torque fluctuations so windings are designed to be as close to single-harmonic as possible. Twice $n_0$ is the polarity of the stator – an intuitive definition because $n_0=1$ produces a dipolar field, $n_0=2$ quadrupolar and so on. The subscript *n* becomes redundant when only one harmonic is considered, so I drop it. An easy transformation converts the 'x-y' representation of stator current to a 'r-θ' notation : $K_{s1} \cos n_0 \vartheta + K_{s2} \sin n_0 \vartheta = K_s \cos n_0 (\vartheta - \alpha)$. Here $K_s$ is the magnitude of $\mathbf{K}_s$ and the angle $\alpha$ runs from 0 to $2\pi/n_0$.

In most studies of motor modeling, the polarity $2n_0$ plays no more significant role than to tweak a few parameter values; hence the modeling can proceed with any arbitrary value of polarity, say 2 or 4 or 24, or with the polarity as a parameter.

In the present problem however, the polarity is of vital importance as it determines whether the rotor can at all be captured inside the stator field or not. The known result is that a dipolar field ($2n_0=2$) cannot trap an object inside it whereas a quadrupolar ($2n_0=4$) or higher order multipolar field can. This is why all papers and patents on magnetic bearings expressly mention the stator polarity as 4 or higher. Let us take the 4-pole stator here, the lowest possible number which can induce trapping of the rotor inside. In generalized cylindrical coordinates $\rho,\theta,z$ the magnetic field of a surface current $(K_0 \cos 2\theta)\hat{\mathbf{z}}$ plastered on a long cylinder has a simple enough form: inside the cylinder it is (since I am interested only in what happens inside the stator)

$$\mathbf{B} \propto (-\rho \sin 2\theta)\hat{\boldsymbol{\rho}} + (-\rho \cos 2\theta)\hat{\boldsymbol{\theta}} \quad , \tag{9}$$

in which specific numerical factors are not important. (This result is derived by writing $\mathbf{B}$ as the curl of the vector potential $\mathbf{A}$, solving a Laplace's equation for $\mathbf{A}$ and employing the appropriate boundary conditions.) In Cartesian coordinates however (9) becomes messy : $x^2$-$y^2$ and $xy$ terms start entering the picture on account of the higher order angular harmonics. On the other hand, for a dipolar current $(K_0 \cos\theta)\hat{\mathbf{z}}$,

$$\mathbf{B} \propto (-\sin\theta)\hat{\boldsymbol{\rho}} + (-\cos\theta)\hat{\boldsymbol{\theta}} \quad , \tag{10}$$

which is expressed trivially in Cartesian coordinates as $\mathbf{B} \propto -\hat{\mathbf{y}}$. Given that for the bearing I will have to do electromagnetism in rotated frames, which are all defined in Cartesian bases related through Cartesian transformations, it will be an enormous convenience if the magnetic field has an easy Cartesian representation. Hence I must understand why it is that a quadrupolar field can act as a trap whereas a dipolar field cannot.

The reason is that a quadrupolar magnetic field has a spatial gradient whereas a dipolar field is uniform. If the quadrupolar rotor is displaced from its position at the stator centre, it will experience some force because of the asymmetry in the stator field. Under the right conditions, this force might be a restoring one. On the other hand, even if a dipolar rotor is displaced, there will be no force on it because of the uniformity of the stator field and it will remain a neutral to its will and matter.

Having established (or at least motivated) this last statement, I make the boldest approximation so far (and in the entire analysis). I replace the quadrupolar (or higher multipolar) stator magnetic field with a perturbed dipolar field, which combines the desirable feature of being unidirectional (and hence allowing an easy Cartesian representation) with the other desirable feature of having a spatial gradient (and hence allowing trapping of objects inside). A perturbation of (10), which does not flout the basic tenet of electromagnetism $\nabla \cdot \mathbf{B} = 0$, is $\mathbf{B} = -f_1(x)\hat{\mathbf{y}}$, for some function $f_1$. It is natural to take $f_1$ as symmetric; further, a concave $f_1$ will be closer to reality than a convex one because the magnetic fields of all multipolar stators increase from centre to periphery. The vector potential which produces this field is readily obtained as $\mathbf{A} = f_0(x)\hat{\mathbf{z}}$, where $f_0$ is an antiderivative of $f_1$.

Because the chosen form of stator field is a perturbation off a dipolar field, it will be generated by a perturbation off a dipolar current i.e. $n_0=1$ in (8). I absorb all these perturbations into the stator winding structure, implying that a current characterized by only two parameters $K_s$ and $\alpha$ sets up the magnetic field containing the perturbation. Equation (10) further implies that the dipolar magnetic field points in a direction rotated clockwise 90º to the line joining the stator current's minimum to its maximum. To summarize the above two paragraphs into an equation, I claim that a stator current $\mathbf{K}_s = K_s \cos(\vartheta-\alpha)\hat{\mathbf{z}}$ produces :

$$\mathbf{A} \propto K_s f_0(x_\alpha)\hat{\mathbf{z}} \quad , \tag{11a}$$

$$\mathbf{B} \propto -K_s f_1(x_\alpha)\hat{\mathbf{y}}_\alpha \quad , \tag{11b}$$

where $x_\alpha,y_\alpha$ is a coordinate basis obtained by rotating $x,y$ through the angle $\alpha$. Since the above magnetic field is primarily dipolar, it is reasonable to expect that the rotor which might be confined by it will be a dipole. Graphical arguments indicate that a displacement of the rotor gives rise to a restoring force if the torque angle is in the second quadrant and a destabilizing force if it is in the first quadrant. Now it is a standard result from motor theory [10] that an induction motor operates in first quadrant in the steady state while a PMSM can be made to operate in any quadrant – this suggests that the synchronous motor might exhibit a confined state if operated in the second quadrant whereas the induction motor might not.

On the basis of this argument I will focus on the analysis of PMSM from now on. Before going over to the rotor modeling, I will introduce definitions of the functions $f_0$, $f_1$ etc. which I will use. The simplest ansatz for concave $f_1$ (and ipso facto the other functions) is

$$f_1(x) = 1 + 4\eta x^2 \quad , \tag{12a}$$

$$f_0(x) = \int^x f_1(x')\,\mathrm{d}x' = x + \frac{4}{3}\eta x^3 \quad , \tag{12b}$$

$$f_2(x) = \frac{d}{dx} f_1(x) = 8\eta x \quad , \tag{12c}$$

where $\eta$ is a positive parameter whose size determines the strength of the gradient in **B**.

Now coming to the rotor, since it is dipolar, its current will be expressible as $K_{r1}\cos\gamma + K_{r2}\sin\gamma$ where $\gamma$ is an angle in the rotor *a,b,c* frame (as $\vartheta$ was in the stator frame). This is why I had claimed while writing the LHS of the mechanical equations that (5) is more than just a cosmetic modification of (4).

Analogous to the stator, I will express the rotor current in magnitude angle terms as $\mathbf{K}_r = \hat{\mathbf{c}}\left[K_r \cos(\gamma - \beta)\right]$. Now for PMSM, the magnitude $K_r$ will be a known fixed quantity $K_{r0}$ as the strength of the embedded permanent magnet is given, and $\beta$ will equal the mechanical angle $\psi$ (upto some constant angle which can be taken to be zero without loss of generality). Hence I can write

$$K_{r1} = K_{r0} \cos\psi \quad , \tag{13a}$$
$$K_{r2} = K_{r0} \sin\psi \quad , \tag{13b}$$

in which $K_{r0}$ depends on the strength of the embedded magnets.

On the stator side I will assume that the magnitude of the stator current $\mathbf{K}_s$ is a constant $K_{s0}$ and that the stator will be excited to produce a constant angle $\delta$ between the rotor magnets and the stator current. Mathematically, $\alpha = \psi + \varphi + \delta$ (note that the angle $\varphi$ has to be included here as precession too changes the orientation of the rotor permanent magnets). Since the rotor torque depends heavily on $\delta$, this angle is known as the torque angle.

This more or less establishes the physics behind the mechanism. To summarize it in one sentence, the confinement of the rotor should be dependent on (*a*) the gradient in the stator magnetic field and (*b*) some function of $\delta$ which changes sign from the first quadrant to the second, likely a cosine. Intuitively we can also expect that the strength of the confinement will also be affected by (*c*) the strength of the stator field and the rotor current. The next Subsection will be more mathematical, devoted to calculating the electromagnetic force and torque vectors which can be plugged in to complete the system (7).

### C. ELECTROMAGNETIC FORCE AND TORQUE

In this Subsection I will calculate the electromagnetic force and torque and thus complete the hanging system (7). To keep the treatment general, and readily extendable to the induction motor, I will work in terms of the stator angle $\alpha$ and the rotor currents $K_{r1}$ and $K_{r2}$ and substitute specific forms only at the end.

The starting expression for electromagnetic force is simple enough :

$$\mathbf{F} = \int i \mathrm{d}\mathbf{l} \times \mathbf{B} \quad , \tag{14}$$

where *i* denotes the wire current and the integral must be made over the entire current configuration. Here since I have surface currents **K** and not wire currents *i*, there will be an additional integration involved. Further, since the force will be generated only through interaction between the rotor current and the stator field (the rotor field cannot interact with the rotor current to give a torque – an isolated electromagnetic body cannot exert a force or torque on itself), the relevant **B** here will be $\mathbf{B}_s$. Incorporating all this I have

$$\begin{aligned}\mathbf{F} &= \int_{-h}^{h} \mathrm{d}c \int_{0}^{2\pi} \mathrm{d}\gamma r_0 \mathbf{K}_r \times \mathbf{B}_s \\ &= \int_{-h}^{h} \mathrm{d}c \int_{0}^{2\pi} \mathrm{d}\gamma r_0 \left[\hat{\mathbf{c}}\left(K_{r1}\cos\gamma + K_{r2}\sin\gamma\right)\right] \times \mathbf{B}_s\end{aligned} \quad . \tag{15}$$

The next step is a calculation of the stator magnetic field $\mathbf{B}_s$. The starting point is the definition (11) :

$$\mathbf{B}_s = -K_{s0} f_1(x_\alpha) \hat{\mathbf{y}}_\alpha \quad . \tag{16}$$

Now I must do a change of basis from *x,y,z* to *a,b,c*; before applying the rotation tensor I have to keep in mind that the rotor CM is displaced from the stator origin. So I must first write

$$X = x - x_{CM} \quad , \tag{17a}$$
$$Y = y - y_{CM} \quad , \tag{17b}$$

and then apply (2) on *X* and *Y*. Using the definition of $x_\alpha$ and $y_\alpha$ and then performing the transformation,

$$\mathbf{B}_s = -K_{s0}\left[-\sin\alpha\{\cos\varphi\hat{\mathbf{a}} - \cos\theta\sin\varphi\hat{\mathbf{b}} + \sin\theta\sin\varphi\hat{\mathbf{c}}\} + \cos\alpha\{\sin\varphi\hat{\mathbf{a}} + \cos\theta\cos\varphi\hat{\mathbf{b}} - \sin\theta\cos\varphi\hat{\mathbf{c}}\}\right] \text{ times}$$
$$\left[f_1\{\cos\alpha(a\cos\varphi - b\cos\theta\sin\varphi + c\sin\theta\sin\varphi + x_{CM}) + \sin\alpha(a\sin\varphi + b\cos\theta\cos\varphi - c\sin\theta\cos\varphi + y_{CM})\}\right] . \quad (18)$$

In equations like this, I have used the word 'times' to indicate multiplication as the usual multiplication symbols (dot and cross) can create confusion when vectors are involved.

I now do three things :

1. I assume the angle of nutation i.e. $\theta$ to be small and the displacements of the CM, $x_{CM}$ and $y_{CM}$, also to be small
2. I separate large and small terms in the argument of $f_1$, do a Taylor expansion and retain only the first nontrivial term
3. I substitute $a = r_0\cos\gamma$ and $b = r_0\sin\gamma$ wherever possible.

The three assumptions, followed by the definitions

$$M = c\theta\sin\varphi + x_{CM} \quad , \quad (19a)$$
$$N = c\theta\cos\varphi - y_{CM} \quad , \quad (19b)$$

and a few routine trigonometric identities reduce (18) to

$$\mathbf{B}_s = -K_{s0}\begin{vmatrix} \hat{\mathbf{a}}\left[\sin(\varphi-\alpha)\{f_1(r_0\overline{\cos\varphi+\gamma-\alpha}) + (M\cos\alpha - N\sin\alpha)f_2(r_0\overline{\cos\varphi+\gamma-\alpha})\}\right] + \\ \hat{\mathbf{b}}\left[\cos(\varphi-\alpha)\{f_1(r_0\overline{\cos\varphi+\gamma-\alpha}) + (M\cos\alpha - N\sin\alpha)f_2(r_0\overline{\cos\varphi+\gamma-\alpha})\}\right] + \\ \hat{\mathbf{c}}\left[-\theta\cos(\varphi-\alpha)f_1(r_0\overline{\cos\varphi+\gamma-\alpha})\right] \end{vmatrix} . \quad (20)$$

Before jumping into the calculation (15) it is worthwhile to get an idea of which terms will survive the integration over $\gamma$. Surviving terms must all have the structure $\cos^2\gamma$, $\sin^2\gamma$, $\cos^2 2\gamma$, $\sin^2 2\gamma$ etc. Since $\mathbf{K}_r$ has only $\cos\gamma$ and $\sin\gamma$ terms, only these harmonics from $\mathbf{B}_s$ will multiply to give a nonzero contribution. The $c$ component of $\mathbf{B}_s$ clearly goes out as it will be crossed into a vector pointing along the $c$ direction and the remainder, after using trigonometric identities and retaining only the dipolar terms, is

$$\mathbf{B}_{s*} = -4\eta r_0 K_s(M\cos\alpha - N\sin\alpha)\begin{vmatrix} \hat{\mathbf{a}}\left[\sin\gamma\{\cos(2\varphi - 2\alpha) - 1\} + \cos\gamma\sin(2\varphi - 2\alpha)\right] + \\ \hat{\mathbf{b}}\left[\cos\gamma\{\cos(2\varphi - 2\alpha) + 1\} - \sin\gamma\sin(2\varphi - 2\alpha)\right] \end{vmatrix} . \quad (21)$$

The notation $\mathbf{B}_{s*}$ indicates that this is not the full $\mathbf{B}_s$ but only a locally relevant form. Taking the cross product and performing the integration over $\gamma$,

$$\mathbf{F}_{abc} = \int_{-h}^{h} dc\ 4\pi\eta r_0^2 K_s(M\cos\alpha - N\sin\alpha)\begin{vmatrix} \hat{\mathbf{a}}\left[K_{r1}\{\cos(2\varphi - 2\alpha) + 1\} - K_{r2}\sin(2\varphi - 2\alpha)\right] - \\ \hat{\mathbf{b}}\left[K_{r1}\sin(2\varphi - 2\alpha) + K_{r2}\{\cos(2\varphi - 2\alpha) - 1\}\right] \end{vmatrix} . \quad (21)$$

One small step remains as $M$ and $N$ are functions of $c$. When these are integrated, they give two 'new' measures of the rotor eccentricity i.e.

$$P = 2hx_{CM} \quad , \quad (23a)$$
$$Q = 2hy_{CM} \quad . \quad (23b)$$

We note that the $\theta$-dependent terms have exited the equation because they are odd functions of $c$ and vanish upon integration over a symmetric interval. Finally, after absorbing the various factors such as height and radius and $\pi$ into a positive constant $\chi_1$, I can write the force as

$$F_a = \chi_1\eta K_s(P\cos\alpha - Q\sin\alpha)\left[K_{r1}\{\cos(2\varphi - 2\alpha) + 1\} - K_{r2}\sin(2\varphi - 2\alpha)\right] \quad , \quad (24a)$$
$$F_b = -\chi_1\eta K_s(P\cos\alpha - Q\sin\alpha)\left[K_{r1}\sin(2\varphi - 2\alpha) + K_{r2}\{\cos(2\varphi - 2\alpha) - 1\}\right] \quad , \quad (24b)$$
$$F_c = 0 \quad . \quad (24c)$$

I can now easily apply the transformation (2) to obtain $\mathbf{F}_{xyz}$ and thus the mechanical equations (7a) and (7b) are completely determined.

An important consistency check is that all the force terms are proportional to the eccentricity variables – the base case of perfectly centred perfectly aligned rotor experiences no force as it should. Now let us turn to the torque vector. The basic equation for torque is $\mathbf{T} = \int \mathbf{r} \times d\mathbf{F}$ with the integral being evaluated all over the rotor periphery; using (15) this becomes

$$\mathbf{T} = \int_{-h}^{h} dc \int_{0}^{2\pi} d\gamma \; r_0 \mathbf{r} \times (\mathbf{K}_r \times \mathbf{B}_s) \quad . \tag{25}$$

From the 'back in a cab' rule,

$$\begin{aligned}\mathbf{T} &= \int_{-h}^{h} dc \int_{0}^{2\pi} d\gamma\, r_0 \left[ \mathbf{K}_r (\mathbf{r} \cdot \mathbf{B}_s) - \mathbf{B}_s (\mathbf{r} \cdot \mathbf{K}_r) \right] \\ &= \int_{-h}^{h} dc \int_{0}^{2\pi} d\gamma\, r_0 \begin{bmatrix} \hat{\mathbf{c}}(K_{r1} \cos\gamma + K_{r2} \sin\gamma) \{(r_0 \cos\gamma \hat{\mathbf{a}} + r_0 \sin\gamma \hat{\mathbf{b}} + c\hat{\mathbf{c}}) \cdot (B_{sa}\hat{\mathbf{a}} + B_{sb}\hat{\mathbf{b}} + B_{sc}\hat{\mathbf{c}})\} \\ -(B_{sa}\hat{\mathbf{a}} + B_{sb}\hat{\mathbf{b}} + B_{sc}\hat{\mathbf{c}})\{(r_0 \cos\gamma \hat{\mathbf{a}} + r_0 \sin\gamma \hat{\mathbf{b}} + c\hat{\mathbf{c}}) \cdot \hat{\mathbf{c}}(K_{r1} \cos\gamma + K_{r2} \sin\gamma)\} \end{bmatrix}\end{aligned} \tag{26}$$

This is a complex expression which I must interpret carefully to avoid unnecessary calculation. The first term in the box bracket is clearly in the $c$ direction while the second term has components in all directions. More tricky is the issue of which multipoles survive the integration over $\gamma$. As usual the survivors will be squares of cosines and sines but they will be generated from more sources than was the case in (15). For instance, looking at the first term in the box, a $\cos 2\gamma$ in $B_{sa}$ might combine with $\cos\gamma$ in $r_a$ (the $a$ component of the $\mathbf{r}$ vector) to produce a $\cos\gamma$ term which can then combine with $K_{r1}\cos\gamma$ to give a nonzero integral. In the second term, the dot product reduces to $K_{r1}\cos\gamma + K_{r2}\sin\gamma$ and the $a$ and $b$ components of torque will be obtained by multiplying the corresponding components of $\mathbf{B}_s$. But now from the multipolar viewpoint I have a situation identical to (15) and again the relevant form of $\mathbf{B}_s$ will be $\mathbf{B}_{s*}$ as found in (23). We note however that the presence of an extra 'c' in this term will cause a difference when the integration over c is carried out. This time, the θ components will be the ones to survive while the xCM and yCM components will exit. Thus, M and N will now turn into two new variables, $P' = (h^3 \theta \sin\varphi)/3$ and $Q' = (h^3 \theta \cos\varphi)/3$. Substituting that into the second term of (28) and doing the integrals exactly as for the force, I have (letting $\chi_2$ be a positive constant which absorbs $\pi$ etc.)

$$T_a = \chi_2 \eta K_s (P'\cos\alpha - Q'\sin\alpha)\left[K_{r1}\sin(2\varphi - 2\alpha) + K_{r2}\{\cos(2\varphi - 2\alpha) - 1\}\right] \quad , \tag{27a}$$

$$T_b = \chi_2 \eta K_s (P'\cos\alpha - Q'\sin\alpha)\left[K_{r1}\{\cos(2\varphi - 2\alpha) + 1\} - K_{r2}\sin(2\varphi - 2\alpha)\right] \quad . \tag{27b}$$

Note that the components of torque which I ignored in (28) are all in $c$ direction, so (29) above is the final form of $T_a$ and $T_b$. They are identical in structure to the expressions for $F_b$ and $F_a$ respectively.

To evaluate the final term $T_c$, it is best (safest) to write out $\mathbf{B}_s$ fully, starting from (21), making all substitutions, assumptions etc. and collecting like multipoles. This process involves nothing other than a few pages of trigonometric identities so I just write the final answer as it is :

$\mathbf{B}_s = -K_s$ times

$$\begin{aligned}&\hat{\mathbf{a}}\begin{vmatrix} (1+2\eta r_0^2)\sin(\varphi-\alpha) + \\ \cos\gamma\left[4\eta r_0(M\cos\alpha - N\sin\alpha)\sin(2\varphi-2\alpha)\right] + \sin\gamma\left[4\eta r_0(M\cos\alpha - N\sin\alpha)\{\cos(2\varphi-2\alpha)-1\}\right] + \\ \cos 2\gamma\left[\eta r_0^2\{\sin(3\varphi-3\alpha) - \sin(\varphi-\alpha)\}\right] + \sin 2\gamma\left[\eta r_0^2\{\cos(3\varphi-3\alpha) - \cos(\varphi-\alpha)\}\right] \end{vmatrix} \\ &+\hat{\mathbf{b}}\begin{vmatrix} (1+2\eta r_0^2)\cos(\varphi-\alpha) + \\ \cos\gamma\left[4\eta r_0(M\cos\alpha - N\sin\alpha)\{\cos(2\varphi-2\alpha)+1\}\right] + \sin\gamma\left[4\eta r_0(M\cos\alpha - N\sin\alpha)\{-\sin(2\varphi-2\alpha)\}\right] + \\ \cos 2\gamma\left[\eta r_0^2\{\cos(3\varphi-3\alpha) + \cos(\varphi-\alpha)\}\right] + \sin 2\gamma\left[\eta r_0^2\{-\sin(3\varphi-3\alpha) - \sin(\varphi-\alpha)\}\right] \end{vmatrix} \\ &+\hat{\mathbf{c}}\begin{vmatrix} -\theta(1+2\eta r_0^2)\cos(\varphi-\alpha) + \\ \cos 2\gamma\left[-\theta\eta r_0^2\{\cos(3\varphi-3\alpha) + \cos(\varphi-\alpha)\}\right] + \sin 2\gamma\left[-\theta\eta r_0^2\{\sin(3\varphi-3\alpha) - \sin(\varphi-\alpha)\}\right] \end{vmatrix}\end{aligned} \tag{28}$$

Now the contribution to $T_c$ which arises from the second term in the box in (28) features multiplication of $B_{sc}$ above with dipolar terms of $\mathbf{K}_r$. Since $B_{sc}$ has no dipoles, this contribution evaluates to zero when the $\gamma$ integration is performed. Hence the only contribution is from the first term in the box. This term is

$$T_c = \int_{-h}^{h} dc \int_{0}^{2\pi} d\gamma r_0 \left( K_{r1} \cos\gamma + K_{r2} \sin\gamma \right) \left( r_0 \cos\gamma B_{sa} + r_0 \sin\gamma B_{sb} + c B_{sc} \right)$$
$$= \frac{r_0}{2} \int_{-h}^{h} dc \int_{0}^{2\pi} d\gamma \begin{bmatrix} K_{r1} B_{sa} \left( 1 + \cos 2\gamma \right) + K_{r1} B_{sb} \sin 2\gamma + c K_{r1} B_{sc} \cos\gamma + \\ K_{r2} B_{sa} \sin 2\gamma + K_{r2} B_{sb} \left( 1 - \cos 2\gamma \right) + c K_{r2} B_{sc} \sin\gamma \end{bmatrix} . \quad (29)$$

Considering each of the six terms in the integrand in the RHS of (31), the harmonics of $\mathbf{B}_s$ which are going to survive in those terms are, respectively, 1 and $\cos 2\gamma$, $\sin 2\gamma$, nothing, $\sin 2\gamma$, 1 and $\cos 2\gamma$, nothing. This finally leads to the value of $T_c$ as (where I absorb geometry etc. into a constant $\chi_3$)

$$T_c = \chi_3 K_{s0} \begin{vmatrix} K_{r1} \left[ \left( 1 + 2\eta r_0^2 \right) \sin(\alpha - \varphi) - \eta r_0^2 \left\{ \sin(3\varphi - 3\alpha) - \sin(\varphi - \alpha) \right\} + \eta r_0^2 \left\{ \sin(3\varphi - 3\alpha) + \sin(\varphi - \alpha) \right\} \right] + \\ K_{r2} \left[ -\left( 1 + 2\eta r_0^2 \right) \cos(\alpha - \varphi) - \eta r_0^2 \left\{ \cos(3\varphi - 3\alpha) - \cos(\varphi - \alpha) \right\} - \eta r_0^2 \left\{ \cos(3\varphi - 3\alpha) - \cos(\varphi - \alpha) \right\} \right] \end{vmatrix} . \quad (30)$$

It is a ready and important check that for the base case of 2D rotation in a uniform dipolar field, I recover the torque formula well known from motor theory [28] : $T = K_{r1} K_{s2} - K_{s1} K_{r2}$.

I am done. The aim of this Section was to find a dynamic model of the system and I have found it. The boxed equations (7), (13), (26), (29) and (32) completely specify the time-evolution of all currents, coordinates and momenta of the three-dimensional magnetic bearing. My next task is the solution of this system; that is kept for the next Section. *attacca*

## 2  Solving the system equations

The key to reducing the formidable equation of motion to a simple and insightful form is a separation of scales analysis. As I have already specified earlier, I will assume a stator control law of $\alpha = \psi + \varphi + \delta$ where $\delta$, a constant, is the torque angle. Now I assume that the spin rate $\dot\psi$ of the rotor about its axis is fast and more or less constant at $\Omega$ while the evolution of all system variables (including $\dot\psi$) is slow. Then I can write $\alpha = \psi + \varphi + \delta = \Omega t + slo$ where $\Omega$ is a constant high frequency and *slo* denotes something whose temporal variation is slow with respect to this frequency scale. This is of course the same argument which is used to solve the Kapitsa pendulum [11,12]. Now note that

$$\cos\alpha = \cos(\Omega t + slo) = \cos\Omega t \cos slo - \sin\Omega t \sin slo \quad , \quad (31a)$$
$$\sin\alpha = \sin(\Omega t + slo) = \sin\Omega t \cos slo + \cos\Omega t \sin slo \quad . \quad (31b)$$

Over a large time scale, i.e. a time scale at which the slow variables operate, the $\cos\Omega t$ and $\sin\Omega t$ terms both average out to zero hence all the expressions in the RHS of (33) are zero. Further, if any slow variable or combination thereof multiplies a $\cos\alpha$ or $\sin\alpha$ term, that too averages out to zero in the long run.

A $\cos^2\alpha$ term or $\sin^2\alpha$ term has a nonzero long time average, for example

$$\cos^2\alpha = \left( \cos\Omega t \cos slo - \sin\Omega t \sin slo \right)^2 = \cos^2\Omega t \cos^2 slo + \sin^2\Omega t \sin^2 slo - 2\cos\Omega t \sin\Omega t \cos slo \sin slo . \quad (32)$$

The $\cos^2\Omega t$ and $\sin^2\Omega t$ average out to 1/2, the $\sin 2\Omega t$ goes to zero and the net average is 1/2. Of course this is the same thing I did a couple pages back with $\Omega t$ here acting as substitute for $\gamma$; in (15) and subsequent material, only squares survived the integral over one whole period, here only squares survive the integral over long time, which is like many whole periods. Based on this analogy, I can give the present process the fancy and meaningless name of time domain multipolar expansion.

I now perform this analysis on (26) and ipso facto on (29) too. The former yields, after substituting $K_{r1}$ and $K_{r2}$ from (13) and the stator control law

$$\mathbf{F}_{abc} = \chi_1 \eta K_{s0} \begin{vmatrix} \hat{\mathbf{a}} \left[ \left( P\cos\alpha - Q\sin\alpha \right) \begin{Bmatrix} K_{r0} \cos(\alpha - \varphi - \delta)(\cos 2\varphi \cos 2\alpha + \sin 2\varphi \sin 2\alpha + 1) - \\ K_{r0} \sin(\alpha - \varphi - \delta)(\sin 2\varphi \cos 2\alpha - \cos 2\varphi \sin 2\alpha) \end{Bmatrix} \right] - \\ \hat{\mathbf{b}} \left[ \left( P\cos\alpha - Q\sin\alpha \right) \begin{Bmatrix} K_{r0} \cos(\alpha - \varphi - \delta)(\sin 2\varphi \cos 2\alpha - \cos 2\varphi \sin 2\alpha) + \\ K_{r0} \sin(\alpha - \varphi - \delta)(\cos 2\varphi \cos 2\alpha - \sin 2\varphi \sin 2\alpha - 1) \end{Bmatrix} \right] \end{vmatrix} . \quad (33)$$

Once again I have a situation where in each component of $\mathbf{F}$ I have temporal dipoles in the first multiplier and a maze of temporal multipoles in the second multiplier. Using trigonometric identities, chucking out irrelevant multipoles and performing the time average, I have

$$F_a = \frac{\chi_1 \eta K_{r0} K_{s0}}{2} \left[ P\left\{ \cos\delta + \cos(\varphi - \delta) \right\} - Q\left\{ \sin\delta + \sin(\varphi - \delta) \right\} \right] \quad , \quad (34a)$$

$$F_b = \frac{\chi_1 \eta K_{r0} K_{s0}}{2} \Big[ P\{\sin\delta + \sin(\varphi-\delta)\} + Q\{\cos\delta + \cos(\varphi-\delta)\} \Big] \;, \tag{34b}$$

$$F_c = 0 \;. \tag{34c}$$

This is quite a neat result out of a cumbersome expression like (35). Of course, for the solution I want $\mathbf{F}_{xyz}$ not $\mathbf{F}_{abc}$, but I will wait for the conversion until after linearizing the system.

The torque components $T_a$ and $T_b$ of course follow from (36) by replacing $\chi_1$ with $\chi_2$ and swapping $a$ and $b$ components. $T_c$ is treated identically to (26) and the final result is

$$T_a = \frac{\chi_2 \eta K_{r0} K_{s0}}{2} \Big[ P'\{\sin\delta + \sin(\varphi-\delta)\} + Q'\{\cos\delta + \cos(\varphi-\delta)\} \Big] \;, \tag{35a}$$

$$T_b = \frac{\chi_2 \eta K_{r0} K_{s0}}{2} \Big[ P'\{\cos\delta + \cos(\varphi-\delta)\} - Q'\{\sin\delta + \sin(\varphi-\delta)\} \Big] \;, \tag{35b}$$

$$T_c = \chi_3 K_{r0} K_{s0} \left(1 + 2\eta r_0^2\right) \sin\delta \;. \tag{35c}$$

Equations (36) and (37) give the RHS of the system dynamics on the slow time scale. As an aside, (37c) makes it clear at once why $\delta$ is called the torque angle.

The last step is linearization of the system. The operating point about which I do this is $\dot\psi = \Omega$ and everything else is zero. I put the $\Delta$ sign explicitly only on $\omega_c$ : for the other variables, I let it be assumed implicitly. A problem which occurs during linearization of any rotational dynamical system about a zero point is that $\omega_b$ becomes equal to $\theta\dot\varphi$ which is second order in small quantities and hence negligible. To work around this problem I must first ensure that the homogeneous part of the system (7c-e), treated as a dynamical system in $\omega_a$, $\omega_b$ and $\omega_c$, is stable. This of course yields the well known Euler stability condition i.e. rotation is stable if it is about the axis with maximum or minimum moment of inertia. This step done, I can now drop $\omega_b$ from the system. The terms $P'$ and $Q'$ simplify as

$$P' = 0 \;, \tag{36a}$$

$$Q' = h^3 \theta / 3 \;. \tag{36b}$$

Since $F_a$ and $F_b$ from (36) are small quantities $(P,Q)$ multiplied by functions of the small quantity $\varphi$, these functions must be evaluated at $\varphi=0$. Then (38) reduces to

$$F_a = P \cos\delta \;, \tag{37a}$$

$$F_b = -Q \cos\delta \;. \tag{37b}$$

Now I must transform them to $\mathbf{F}_{xyz}$ using (2); once again all small quantities must be set to zero and

$$F_x = P \cos\delta \;, \tag{38a}$$

$$F_y = -Q \cos\delta \;. \tag{38b}$$

$T_a$ and $T_b$ of course have the same forms as $F_b$ and $F_a$, and $T_c$ does not change from its form (37c).

Finally I have all the components of the linearized system. There are seven significant variables here; out of the original ten, $\omega_b$ drops out as it is too small, $\varphi$ and $\psi$ coalesce since for small $\theta$ they tend to the same angle, and there is only one equation which describes the evolution of their combined derivative i.e. $\omega_c$. Recognizing that $\omega_a = \dot\theta$, I finally rewrite (7) in linearized form together with the corresponding forms of the RHSes. Factoring in that the torque $T_c$ in (37c) goes into maintaining the basic speed $\omega_c = \Omega$, I have

$$m\ddot{x}_{CM} + \kappa_1 \dot{x}_{CM} \left(-2h\chi_1 \eta K_{r0} K_{s0} \cos\delta\right) x_{CM} = 0 \;, \tag{39a}$$

$$m\ddot{y}_{CM} + \kappa_1 \dot{y}_{CM} + \left(-2h\chi_1 \eta K_{r0} K_{s0} \cos\delta\right) y_{CM} = 0 \;, \tag{39b}$$

$$I\ddot\theta + \kappa_2 \dot\theta + \left(-h^2 \chi_2 \eta K_{r0} K_{s0} \cos\delta\right) \theta = 0 \;, \tag{39c}$$

$$I_c \Delta\dot\omega_c + \kappa_2 \Delta\omega_c = 0 \;. \tag{39d}$$

This is the final equation of motion. Some features and limitations are apparent; I discuss them below.

The primary feature of the linearized equations is that they describe a damped harmonic oscillator or damped harmonic repeller (exponential solutions) depending on the signs of the coefficients. The oscillatory solutions occur if $\cos\delta<0$. Prima facie, the rotor has been confined inside the stator. Unfortunately there is a caveat which has forced me to introduce the word 'basic' in the title of this work. And that caveat is not Earnshaw's theorem – see the next paragraph for that. It is that two of the spring constants (the ones describing the $x$ and $y$ motions) are identical. Considering the stiffness matrix for these two degrees of freedom only, it means that this matrix is diagonal with the same non-zero element appearing in both positions. Now this analysis is obviously approximate – I replaced the actual magnetic field with a tractable equivalent.

Even a more refined analysis using a better field will also be approximate at some level – a calculation with a genuinely real magnetic field is impossible. And the effect of the ignored terms might well be to introduce a coupling between the *x* and *y* degrees of freedom. In such a case, the addition of even the smallest off-diagonal terms to the stiffness matrix might cause its eigenvalues to acquire imaginary parts, thereby robbing the matrix of its positive-definiteness. Thus, although the gradient in **B** and the second quadrant torque angle amount to a stably confined state, the stability is tissue-thin. Further, I have not discussed the issue of how the rotor weight will be supported. If the setup is mounted horizontally, the only forces to balance the weight will arise from the gradient of the stator field and not the stator field itself. This is a small force and cannot balance a practical rotor. If the setup is mounted vertically and a separate field configuration used for *z*-confinement then those fields must be quite strong, and care must be taken to ensure that they do not destabilize the rotor. Thus this analysis, although leading to some conditions necessary for confinement of the rotor, is partial, and for designing a realistic levitating motor, a more complex and intricate magnetic field will be required. I will leave these pressing questions for subsequent works, bringing the present Section to a close with a comment on the applicability of Earnshaw's theorem.

The question may arise that, assuming (41) to be true and assuming also that some mechanism to confine the rotor in *z* is indeed found and installed, does the absolute confinement of the rotor not amount to a violation of Earnshaw's theorem. The answer to this is fortunately no, because the fields involved in trapping the rotor are time dependent. A constant second quadrant torque angle necessarily means that the applied fields are changing in time, and Earnshaw's theorem does not apply. In the steady state, when the speed of rotation is constant, you might argue that the situation is essentially static in a frame rotating synchronously with the rotor – that is true but then there is the extra force of air drag which is keeping the speed constant, and Earnshaw's theorem does not hold in the presence of drag. If I really made the external field static (supplied the stator with dc) then the rotor would work its way into a first quadrant torque angle position and then get thrown out laterally, preserving Earnshaw's theorem. The constant change in field position, made possible through control, circumvents the theorem, just as controlled motion applied to the base of a pendulum stabilizes the statically unstable inverted position.

In view of the theoretical nature of this work, I will now also include a Subsection on dynamic modeling of induction rotors. Although the induction rotor is an unsuitable confinement candidate, the following procedure is a very general method for obtaining the dynamic model of any eddy current structure, and might thus be applied to the analysis of eddy current based bearings.

## 3 Eddy current dynamic model

First I will present the philosophy behind the electrodynamic modeling. The overall process is a satisfaction of consistency of the voltage between top and bottom of the rotor when measured in two different ways. One way is direct. Suppose the rotor carries a surface current along its axis ($\hat{\mathbf{c}}$) of magnitude $K_r$ and angle $\beta$ (both whose dynamics we want to find). Then, assuming the rotor material to be linear, the voltage between top and bottom must be proportional to $K_r \cos(\gamma - \beta)$ through some constant ("const."). The second way of measuring voltage is roundabout. Since the rotor is carrying a time varying current $\mathbf{K}_r$, that will create a time varying vector potential $\mathbf{A}_r$ which in turn will induce an electric field $-\partial \mathbf{A}_r / \partial t$ in the rotor, from Lenz' law. Further, $\mathbf{K}_r$ will also create a magnetic field $\mathbf{B}_r$, and since the rotor is moving through this field (my reference frame is stationary with respect to the *stator*, or the lab) there will be a $\mathbf{v} \times \mathbf{B}_r$ term. More electric fields will be induced in the rotor on account of the stator magnetic field : these will be $-\partial \mathbf{A}_s / \partial t$ and $\mathbf{v} \times \mathbf{B}_s$. When the resultant of all these electric fields are integrated over the rotor height, that will give a voltage. To this must be added the voltage, if any, which I am applying on the rotor from outside. Since the IM rotor is a short circuited structure, this last term is zero. Now the voltage obtained from the direct way must equal that from the roundabout way and I have

$$\text{const} \cdot \mathbf{K}_r = -\frac{\partial \mathbf{A}_r}{\partial t} + \mathbf{v} \times \mathbf{B}_r - \frac{\partial \mathbf{A}_s}{\partial t} + \mathbf{v} \times \mathbf{B}_s + \cancel{V_{\text{outside}}} \quad , \tag{40}$$

in which I have absorbed the effects of integration into const. Now $\mathbf{A}_r$ and $\mathbf{B}_r$ are functions of $K_r$ and $\beta$ while $\mathbf{A}_s$ and $\mathbf{B}_s$ are functions of $K_s$ and $\alpha$ (recall that these two are known quantities); writing these dependences explicitly and rearranging terms I have

$$\frac{\partial}{\partial t} \mathbf{A}(K_r, \beta) + \text{const} \cdot (K_r, \beta) - \mathbf{v} \times \mathbf{B}(K_r, \beta) = -\frac{\partial}{\partial t} \mathbf{A}(K_s, \alpha) + \mathbf{v} \times \mathbf{B}(K_s, \alpha) \quad . \tag{41}$$

Implicit in this equation is the magnetostatic approximation – had that not been assumed then extra terms featuring time derivatives of currents would have had to be included in the expressions for magnetic field. But motors always operate in the region where this is valid (a typical speed in a motor is 100 m/s, orders of magnitude lower than the speed of light) hence (14) is exact for all practical purposes. Thus, a structure of the eddy current dynamic model has emerged and it now remains to calculate each term to the necessary level of accuracy. Note that this structure makes no assumptions about the rotor geometry and uses principles which are universally true for electromagnetic systems; thus it can be used (mutatis mutandis) to analyse eddy currents in any given system.

Although $\mathbf{A}_r$, $\mathbf{B}_r$, $\mathbf{A}_s$ and $\mathbf{B}_s$ are generally very complex, simplifications generally ensue from a multipolar expansion, wherein contributions other than the fundamental harmonic can usually be thrown away. At this point, I can stop this discussion since it is useless to apply it to the present configuration. We hope that enough information has been conveyed for the procedure to be applicable to configurations where the outcome will have practical utility.

In conclusion I would like to highlight that this is primarily a theoretical Article, describing a method of accurately analysing the dynamics of any given magnetic bearing. Although a practically feasible bearing topology has not been demonstrated, some necessary stability conditions have been found, which can act as a starting point for a more realistic design. The analysis of that design can be performed using the principles developed here, and such a task is currently being reserved for the future works.

<div style="text-align:center">✻   ✻   ✻   ✻   ✻</div>